\documentclass{article}
\usepackage{spconf,amsmath}
\usepackage{amssymb,latexsym, fancyhdr, epsfig}

\title{Information Measures for Microphone Arrays}
\name{Mohamed F. Mansour}
\address{Amazon Inc., USA}
\begin{document}
%
\maketitle
\begin{abstract}
We propose a novel information-theoretic approach for evaluating microphone arrays that relies on the array physics and geometry rather than the underlying beamforming algorithm. The analogy between Multiple-Input-Multiple-Output (MIMO) wireless communication channel and the acoustic channel of microphone arrays is exploited to define information measures of microphone arrays, which provide upper bounds of the information rate of the microphone array system.
\end{abstract}
\section{Introduction}
\label{sec:intro}
Microphone arrays have become an increasingly popular technology for acoustic front-end systems due to its superior speech enhancement performance when compared to single microphone systems. The decreasing hardware cost enables their deployment in mainstream consumer electronics products, e.g., mobile phones and smart speakers. The diversity provided by the microphone arrays is exploited to improve the signal-to-noise ratio and reduce room reverberation for better end-user experience. Further, it is an enabling technology for many signal processing algorithms, e.g., source localization and sound source separation. 

The design of microphone arrays involves different parameters, e.g., number of microphones, overall microphone array area, array geometry, and surface properties. In commercial systems, hardware/software costs are the deciding factors for the number of microphones in the system. Then, for a given array size the design objective depends on the end system. For example, in voice communication systems the system metric is usually subjective or objective speech quality measures, e.g.,  \cite{mushra, pesq, polqa}. In other scenarios, the back-end system is a personal assistance system based on large-vocabulary continuous speech recognition system, e.g., \cite{sarikaya2017technology}; and the more appropriate design criterion is Word Error Rate (WER) and/or False Rejection Rate (FRR) of a system keyword. In the microphone array literature, there are few design criteria that were usually used to evaluate its performance, e.g., beam pattern, directivity index, and white-noise gain \cite{MA_book, benesty2008microphone}. The design of microphone arrays has been traditionally done through heuristics that link the microphone array metrics to the overall system metrics, and solving the resulting optimization problem \cite{gazor1995criteria, ward1995theory, kodrasi2011microphone, feng2012placement, berger1991microphone, markovich2013performance}. 

This traditional design approach has few issues:
\begin{enumerate}
\item The performance depends  on the underlying beamforming algorithm rather than the microphone array physics.
\item It assumes perfect knowledge of the source location, which is not usually achievable.
\item It is not directly linked the overall system objective, therefore, it relies on heuristics that do not always hold.
\item It cannot be extended to the case of multiple sources.
\end{enumerate}
A notably different approach for microphone array analysis that is physics-based was proposed in \cite{svd},  where the singular vectors of the infinite-dimensional singular value decomposition (SVD) of the steering vectors matrix is computed. This metric is independent of the beamforming algorithm but does not resolve the other issues. 

To remedy these issues, we propose a novel information-theoretic approach for evaluating microphone arrays. The approach utilizes the channel capacity concept in information theory literature \cite{shannon}, where the microphone array system is modeled as a Single-Input-Multiple-Output (SIMO) communication channel and its channel capacity is the microphone array metric.  This metric measures the amount of information that could be communicated reliably through the microphone array channel, which is directly related to the overall system objectives. It is independent of the underlying beamforming algorithm. Rather, it is solely determined by the physics of the system and the underlying noise model.  The noise model could be straightforwardly combined with an interference model using the available results from wireless communication literature \cite{MIMO_cap}. Further, a model with unknown speaker position is straightforwardly mapped to a model with imperfect channel knowledge at the receiver. Other generalizations are also discussed where we show few examples that establish the effectiveness of the proposed metric. 

The following notations are used throughout the paper. A bold lower-case letter denotes a column vector, while a bold upper-case letter denotes a matrix. ${\bf{A}}^\prime$ denotes the conjugate transpose of $\bf{A}$, ${\bf{A}}^T$ denotes the transpose of  $\bf{A}$, and ${\bf{A}}_{m,n}$ is the matrix entry at position $(m,n)$. ${\bf{d}}_m$ denotes the $m$-th entry of the vector $\bf{d}$. $\boldsymbol{\theta} \triangleq \left(\theta_a, \phi \right)^T$ denotes the azimuth and elevation angles, respectively, in a spherical coordinate system. $\mathbb{E} \{.\}$ denotes the expectation operator. $M$ always refers to the number of microphones. Additional notations are introduced when needed.

\section{Background}
Shannon described in his seminal $1948$ paper \cite{shannon} the concept of channel capacity that defines the upper bound of information rate that could be communicated reliably across a communication channel. The channel capacity is defined as:
\begin{equation}
C = \text{max}_{p(x)} I(X,Y)
\end{equation}
where $I(X,Y)$ is the mutual information between $X$ (the channel input) and $Y$ (the channel output). The maximization is over all possible distribution of the input variable. The Additive White Gaussian Noise (AWGN) channel of bandwidth B Hz has form
\begin{equation}
Y =  X + Z
\end{equation}
where $ Z \sim \mathcal{N}(0,\sigma^2)$, and its channel capacity in bits per second is \cite{cover}
\begin{equation}
C = B \log_2\left(1+\frac{S}{\sigma^2}\right)  \label{eq_Cfull}
\end{equation}
where $S$ is the signal energy and $S/\sigma^2$ is the SNR. If the bandwidth is B Hz, and the noise PSD is flat at $N_o$, then, $\sigma^2=BN_o$, and the channel capacity in bits/sec/Hz could be expressed from \eqref{eq_Cfull} as
\begin{equation}
C_{AWGN} = \log_2 \left(1 + \text{SNR}\right)
\end{equation}
Now consider the Single-Input Multiple-Output (SIMO) of the form
\begin{equation}
y_l(t) = h_l x(t) + n_l(t) \ \ \ \ \ \ l = 1, 2, ... , L   \label{eq_SIMO_model}
\end{equation}
where $h_l$ is a complex variables, and $n_l(t) \sim \mathcal{CN}(0,\sigma^2 = N_oB)$ (iid complex Gaussian), with the independent noise component at  different channels (i.e., with sample covariance matrix  $\boldsymbol{\Gamma}(t) = \sigma^2\bf{I}_L$ ). For notational convenience, we use the vector form
\begin{equation}
{\bf{y}}(t) = {\bf{h}} \ x(t) + {\bf{n}}(t)    \label{eq_SIMO_vector}
\end{equation}
If ${\bf{h}}$ are perfectly known at the receiver, then the channel capacity in bits/sec/Hz has the form \cite{Tse}
\begin{equation}
C_{SIMO} = \log_2\left(1+ \frac{P \|{\bf{h}}\|^2}{\sigma^2} \right)
\end{equation}
where $P$ is the average transmitted power.
If the noise process is not iid, then the covariance matrix $\boldsymbol{\Gamma}$ is not diagonal.  $\boldsymbol{\Gamma}$  is positive semi-definite with an SVD of the form
\begin{equation}
\boldsymbol{\Gamma} = \bf{U}\bf{S}\bf{U}^\prime \label{Gamma_svd}
\end{equation}
where $\bf{S}$ is a diagonal matrix with the singular values of $\boldsymbol{\Gamma} $ on its diagonal, and $\bf{U}$ is a unitary matrix.
Then consider the following transformation
\begin{equation}
\tilde{\bf{y}} (t) \triangleq {\bf{S}}^{-\frac{1}{2}} \bf{U}^\prime {\bf{y}}(t) = \tilde{\bf{h}} \ x(t) + \tilde{\bf{n}}(t)  \label{eq_simo_white}
\end{equation}
where 
\begin{eqnarray}
\tilde{\bf{h}} &\triangleq& {\bf{S}}^{-\frac{1}{2}} \bf{U}^\prime {\bf{h}} \label{htilde} \\
\tilde{\bf{n}}(t) &\triangleq &{\bf{S}}^{-\frac{1}{2}} \bf{U}^\prime {\bf{n}}(t) 
\end{eqnarray}
The covariance matrix of $\tilde{\bf{n}}(t)$ becomes $\bf{I}$.  The transformation in \eqref{eq_simo_white} does not change the information content as long as ${\bf{S}}$ is full-rank. Hence, for coherent noise we could use the whitened form in \eqref{eq_simo_white} to compute the channel capacity rather than the general representation in \eqref{eq_SIMO_vector}. In this case the channel capacity has the form:
\begin{equation}
C_{SIMO-C} = \log\left(1+ P \|\tilde{\bf{h}}\|^2\right)
\end{equation}
where $\tilde{\bf{h}}$ is as defined in \eqref{htilde}.

Note that, the above discussion made two simplifying assumptions:
\begin{enumerate}
\item The receiver has perfect knowledge of the channel, $\bf{h}$.
\item The channel has only a single user. 
\end{enumerate}
Models with these requirements relaxed are available at the literature, e.g., \cite{MIMO_cap}.


\section{Microphone Array with Free-Space Propagation}

The narrowband far-field free space propagation model for a microphone array at time $t$, frequency $f$, and incidence angle $\boldsymbol{\theta}$ has the general form
\begin{equation}
{\bf{y}}(t,f,\boldsymbol{\theta}) = {\bf{d}}(f,\boldsymbol{\theta}) \ x(t,f) + {\bf{w}}(t,f)    \label{eq_ma_model}
\end{equation}
where ${\bf{d}}(f, \boldsymbol{\theta})$ is the steering vector at frequency $f$ and incidence angle $\boldsymbol{\theta}$ (which includes both azimuth and elevation), $x(t,f)$ is the source signal, and ${\bf{w}}(t,f)$ is the spatial noise signal. In the far-field case, the steering vector has the form
\begin{equation}
 {\bf{d}}(f,\boldsymbol{\theta}) =  \left(e^{-j2\pi f \tau_1(\boldsymbol{\theta}) }  \ \hdots \ \ e^{-j2\pi f \tau_M(\boldsymbol{\theta}) } \right)^T
\end{equation}
where $\tau_k(\boldsymbol{\theta})$ is the time-delay at the $k$-th microphone for a plane-wave with incidence angle $\boldsymbol{\theta}$. In the near-field case, the steering vector for a point source at spherical coordinate $(r,\boldsymbol{\theta})$ is
\begin{equation}
{\bf{d}}(f,r,\boldsymbol{\theta}) = \left(\alpha_1(r) e^{-j2\pi f \tau_1(r,\boldsymbol{\theta}) }  \ \hdots \ \ \alpha_M(r)e^{-j2\pi f \tau_M(r,\boldsymbol{\theta}) } \right)^T \label{NF_steering}
\end{equation}
where $\{\alpha_k(r)\}$ are the attenuations of the wavefront, which is inversely proportional with the distance between the source and the microphone array.  
The covariance matrix of the spatial noise is $\boldsymbol{\Gamma} (t,f) \triangleq \mathbb{E} \{ {\bf{w}}(t,f) {\bf{w}}^\prime(t,f)\}$. If the spatial noise power is distributed as $\sigma_{\bf{w}}^2 (f,\boldsymbol{\theta})$, then \cite{MA_book}
\begin{equation}
\boldsymbol{\Gamma}_{m,n} (f) = \int_0^{2\pi}\int_0^\pi {\bf{d}}_m(f,\boldsymbol{\theta}){\bf{d}}_n^\prime(f,\boldsymbol{\theta}) \sigma_{\bf{w}}^2 (f,\boldsymbol{\theta}) \sin\theta \ d\theta d\phi
\end{equation}
The spatial noise is traditionally modeled as either spherical  or cylindrical diffuse noise; and the corresponding covariance matrix is known \cite{MA_book}. For example, for spherical diffuse noise, the covariance matrix at frequency $f$ has the form 
\begin{equation}
\boldsymbol{\Gamma}_{m,n} (f) = \left\{\begin{array}{rcl} 
\sigma^2  \ \ \ \ \  \ \ \ \ \   \ \ \ \ \   \ \ \ \ \   \ \ \ \ \  \mbox{if } m = n \\
\sigma^2 \text{sinc}\left(2\pi f l_{mn}/c\right)/(1+ \epsilon) \ \ \ \ \  \mbox{otherwise}
\end{array} \right.  \label{Gamma_def}
\end{equation}
where $c$ is the speed of sound, $l_{mn}$ is the distance between microphones $m$ and $n$, and $\epsilon$ is the relative incoherent noise component at microphone $m$. 

\begin{figure}[h]
\includegraphics[width=8cm, height=1.3cm,trim=30mm 25mm 45mm 20mm,clip]{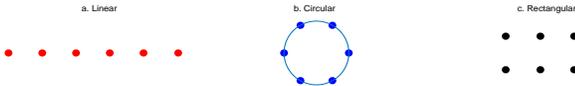}
\vspace{-0.3cm}
\caption{Microphone Array Examples}
\label{fig:MA_example}
\end{figure}

The analogy between the free-space propagation model in \eqref{eq_ma_model} and the general SIMO model in \eqref{eq_SIMO_vector} is obvious. Hence, the channel capacity results from the previous section could be straightforwardly applied to the narrowband free-space propagation model to compute $C(f,\boldsymbol{\theta})$ as 
\begin{equation}
C(f,\boldsymbol{\theta}) =  \log\left( 1+ 2 \|{\bf{S}}^{-\frac{1}{2}} {\bf{U}}^\prime {\bf{d}}(f,\theta) \|^2 \right)
\end{equation}
where $\bf{U}$ and $\bf{S}$ are from the SVD of $\boldsymbol{\Gamma}(f)$ as in \eqref{Gamma_svd}. As an example, consider the three microphone array configurations of Fig. \ref{fig:MA_example}. The three configurations have the same spacing of $3$ cm between adjacent microphones.  

\begin{figure}[h]
\includegraphics[width=8.8cm, height=4.7cm,trim=0mm 0mm 0mm 0mm,clip]{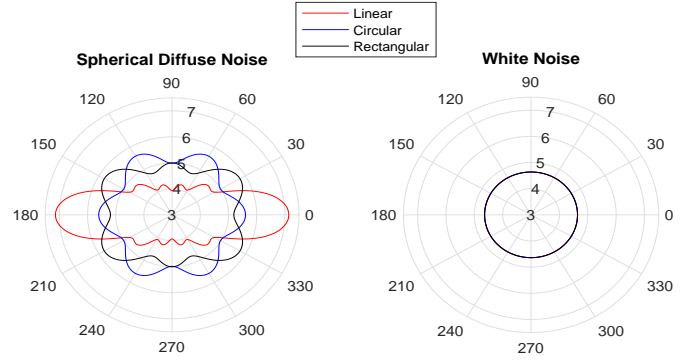}
\vspace{-0.1cm}
\caption{$C(f = 1\text{ kHz}, \theta_a, \phi = \pi/2)$ at different azimuth angles for the microphone array examples}
\label{fig:ex1}
\end{figure}

\begin{figure}[h]
\includegraphics[width=9.7cm,height=4.5cm, ,trim=21mm 00mm 15mm 0mm,clip]{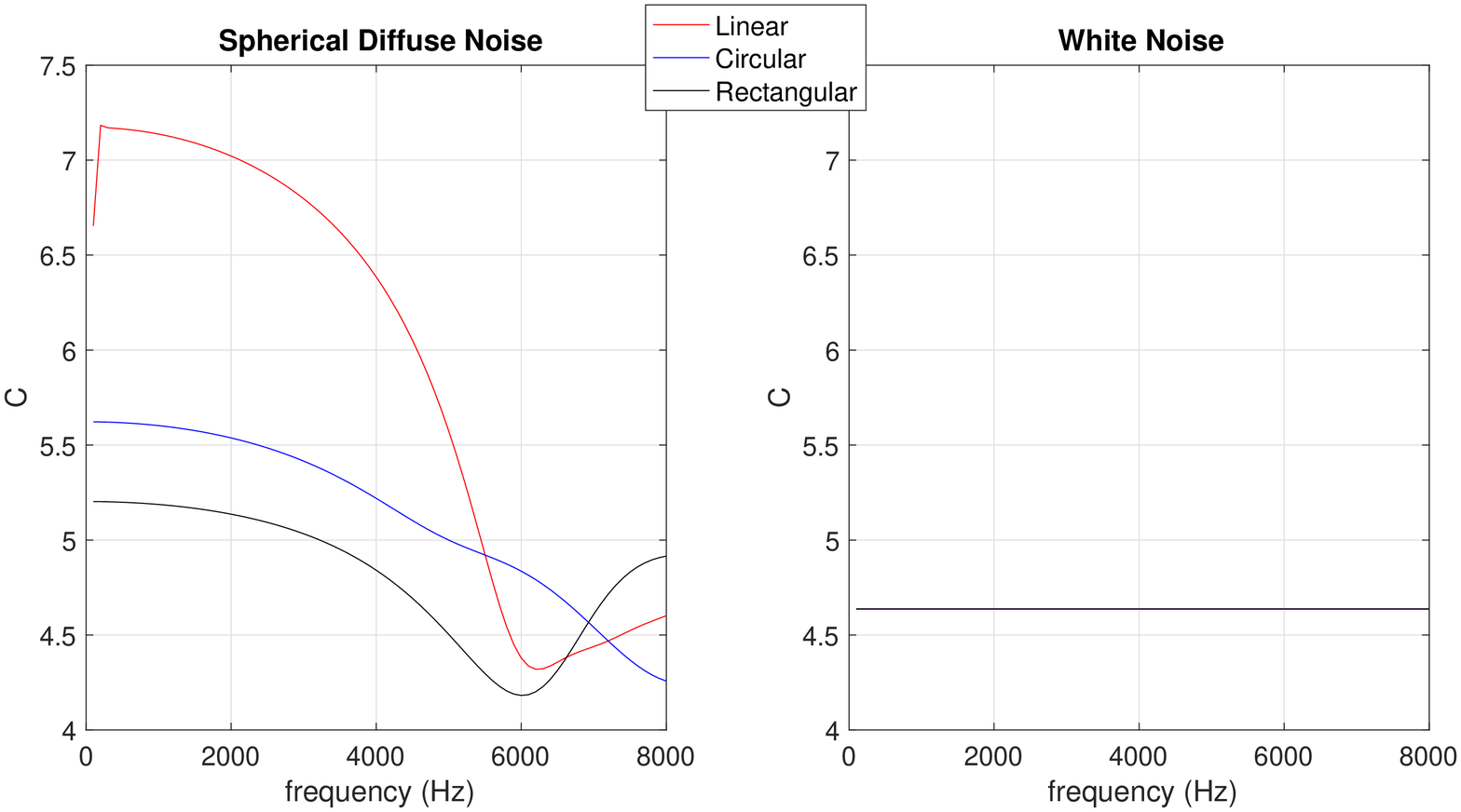}
\caption{$C(f, \boldsymbol{\theta} = [0 \ \ \pi/2])$ for the microphone array examples}
\label{fig:ex2}
\end{figure}

In Fig. \ref{fig:ex1}, we show the channel capacity of the different microphone arrays at $f = 1$ kHz. We show the channel capacity at the horizontal plane (i.e., $\phi = \pi/2$) with both spherical diffuse noise and incoherent noise at SNR = $6$ dB. 
This simple example highlights the effectiveness of the proposed metric to capture the microphone array properties: 
\begin{enumerate}
\item The shape of the microphone array does not matter if the noise is white. In this case, the best receiver is the delay-and-sum beamformer.
\item The linear array is significantly skewed towards the x-axis, while the performance is  compromised along the y-axis. The rectangular array is also skewed towards the x-axis but with less variance.
\item The circular array shows symmetric behavior versus $\theta_a$, with the $6$ cycles in the range $[0, 2\pi]$ that matches the number of microphones.
\end{enumerate} 
A similar analysis with frequency as the independent variable is shown in Fig. \ref{fig:ex2}. Similarly, analysis under various noise/array conditions could be readily performed in a similar way.

\begin{figure}[h]
\includegraphics[width=8.7cm, height=4.2cm,  ,trim=1mm 5mm 0mm 0mm,clip]{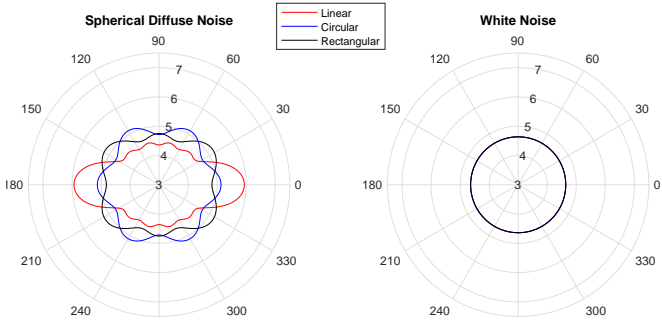}
\caption{$\bar{C}(\boldsymbol{\theta} = [0 \ \ \pi/2])$ for the microphone array examples}
\label{fig:ex3}
\end{figure}

For a broadband signal, the channel capacity needs to be \emph{aggregated} over the frequency range of interest. Therefore, at each incidence angle the average capacity becomes:
\begin{equation}
\bar{C}(\boldsymbol{\theta}) = \mathbb{E}_f \{ C(f,\boldsymbol{\theta})\} \label{sys_metric}
\end{equation}
where $\mathbb{E}_f \{.\}$ is the expectation operator over frequency. In the simplest case, this is reduced to the sample mean of the capacity across all frequencies. In other cases, e.g., for a speech signal it could be a windowed mean with a window function that reflects a typical speech spectrum.  The broadband capacity for $\boldsymbol{\theta} = (0, \pi/2)^T$ is shown in Fig. \ref{fig:ex3}.

The channel capacity concept has proven to be fundamental in communication because it maps naturally to standard metrics in communication systems, e.g., the bitrate. The concept has been applied to other contexts, e.g., audio coding \cite{JJ}. Further, the input-output mutual information is directly correlated with the achievable Minimum Mean Square Error (MMSE) for a general class of channels \cite{verdu1}. This property renders the proposed channel capacity metric as a natural tool for evaluating  microphone arrays, when the overall objective function is dependent on MSE, e.g., for voice communication. In fact, it is straightforward to demonstrate that the MSE after the optimal multichannel Wiener filter \cite{benesty2008microphone} has the same shape as the proposed metric. Nevertheless, a key advantage of the proposed metric is that it could be straightforwardly extended to many other practical usage cases. 

\section{Generalizations}
In the following, we briefly describe various generalizations to the baseline model presented in the previous section. Full details are provided in an expanded version of this work.
\begin{itemize}
\item \emph{Near-Field vs. Far-Field}: In the near-field case the steering vector in \eqref{NF_steering} is used. The system metric in \eqref{sys_metric} would then be $\bar{C}(r,\boldsymbol{\theta})$.
\item \emph{Device Scattering}:
The channel capacity is related to the incidence angle only through the steering vector $\bf{d}$ of the corresponding plane-wave in \eqref{eq_ma_model}. This corresponds to the total wavefield observed at the microphone array, which has the general form
\begin{equation}
{\bf{d}}(f, \boldsymbol{\theta}) = {\bf{d}}^{(I)}(f, \boldsymbol{\theta}) + {\bf{d}}^{(S)}(f, \boldsymbol{\theta})
\end{equation}
where ${\bf{d}}^{(I)}$ and ${\bf{d}}^{(S)}$ refer, respectively, to the incident and scattered wavefield. In a free-space model,  ${\bf{d}}^{(S)}$, which accounts for the impact of the surface of the microphone array, is ignored. In a practical setting, it is computed either using simulation, e.g., through finite-element analysis of the acoustic wave equation, or using physical anechoic measurement. In either case, the resulting response becomes steering vector in \eqref{eq_ma_model}, and the remaining analysis is the same.
\item \emph{Interference}: The proposed model can straightforwardly handle the the presence of one or more point source interferers following the model for MIMO capacity in the presence of interference \cite{blum2003mimo}. If the interferer position is known, then the impact on the narrowband channel capacity is modeled as an additive component to $\boldsymbol{\Gamma}$, where 
\begin{equation}
\boldsymbol{\Gamma}_{m,n}^{(I)}(f) = \boldsymbol{\Gamma}_{m,n}(f) +  {\bf{d}}_I(f){\bf{d}}_I^H(f)
\end{equation} 
where $\boldsymbol{\Gamma}_{m,n}(f)$ is as in \eqref{Gamma_def}, and ${\bf{d}}_I(f)$ is the interferer steering vector at the microphones. 
\item \emph{Unknown Speaker Position}: In practical scenarios, the location of the speaker is usually unknown and the system needs to estimate the correct position prior to passing the beamformed audio to backend processing. The general procedure is to treat the source localization error as an interference component, and average its impact by an the statistics of the localization error. 
\item \emph{Multiple Speakers }: The multiple speakers resembles the multiple access MIMO system \cite{MIMO_cap}. In this case, the channel model becomes MIMO rather than SIMO and spatial diversity is exploited to allow  processing of simultaneous signal sources. 
\end{itemize}
In general, exploiting the resemblance to MIMO wireless channel is a key advantage of the proposed metric as it enables the study of many other cases using the available rich literature on the subject. 


\section{Conclusion}
The proposed information-theoretic metric provides a new direction for effectively evaluating microphone arrays, which would have a significant impact of the overall development cost. Its key advantages over earlier approaches are:
\begin{enumerate}
\item  It is solely determined by the underlying physics and it is independent of the beamforming algorithm.
\item Most practical use cases of microphone arrays resemble similar cases in MIMO wireless system. Hence, the available results from wireless communication literature can be readily applied to the microphone array case.
\end{enumerate}

We showed few design examples that establish the effectiveness of the proposed method, and provided general description for tackling other practical scenarios, e.g., near-field beamforming, and unknown speaker position. Further, the metric is applicable beyond planar microphone arrays by exploiting the surface scattering model for generalized steering vectors.

\bibliographystyle{IEEEbib}
\bibliography{refs}

\end{document}